# Leaf electronics: Nature-based substrates and electrodes for organic electronic applications

Rakesh Rajendran Nair[1], Laura Teuerle[1], Jakob Wolansky[1], Hans Kleemann[1], and Karl Leo[1]

[1]: Dresden Integrated Center for Applied Physics and Photonic Materials (IAPP) and Institute of Applied Physics, Technische Universität Dresden, Nothnitzer Str. 61, 01187 Dresden, Germany

rakesh_rajendran.nair@tu-dresden.de

*Abstract*— The need to reduce the environmental impact of inorganic electronic systems is pressing. Although the field of organic electronics provides a potential solution to this issue, research and optimization is still majorly carried out on glass or plastic substrates. Additionally, the fabrication of organic devices requiring transparent electrodes is fraught with complex techniques and expensive materials which limit widespread implementation and sustainability goals. Here, we show that the quasi-fractal lignocellulose structures extracted from natural leaves can be successfully modified to be used as biodegradable substrates as well as electrodes for optoelectronic applications. Chemically coating the microstructures of these leaf skeletons with metals results in quasi-transparent, flexible electrodes having sheet resistances below 1 $\Omega/\square$ and a concomitant current carrying capacity as high as 6 A over a 2.5 × 2.5 cm² leaf electrode, all while maintaining broadband optical transmittance values of around 80%.

## I. Introduction

Organic electronics is a field poised for ubiquitous implementation in broad walks of life. It is also one toted to be much more environmentally friendly compared to current, commercial inorganic electronics. However, some pertinent issues remain. Foremost is the fact that most research into organic device fabrication is currently being carried out on glass or plastic substrates which are energy-intensive in terms of recycling. For a truly environmentally friendly organics domain, more research efforts need to be directed towards the development and optimization of devices on inexpensive and environmentally friendly/biodegradable substrates. Another issue is the need for transparent or quasi-transparent electrodes. Currently, the fabrication of highly conducting layers with high light transmittance is rife with processing and cost related issues that are slowly being mitigated with research [1].

Here we propose a solution based on naturally occurring tree leaves which may allow biomaterials based on lignocellulose to offer a potential solution to these two bottlenecks while maintaining the environmental sustainability goal that underlies organic electronics research.

Unlike biodegradable polymers such as poly(glycolic acid) (PGA), poly(lactic acid) (PLA) etc. [2], or biodegradable substrates such as paper [3], biopolymers are naturally occurring and are produced by living organisms. Among such materials, cellulose, and lignin (commonly found together and hence termed 'lignocellulose') constitute the most abundant biopolymers on Earth [4] [5] and in combination with hemicellulose, they constitute the major components of a leaf's vasculature. These porous, quasi-fractal structures when combined further with non-toxic, biocompatible and biodegradable polymers such as ethyl cellulose [6] [7] [8] for example, can result in flexible sheets implementable as substrates.

Lignocellulosic polymers if effectively implemented as replacements of pollution-causing units of mass-produced technology (such as substrates) can provide an advantage within the purview of circular economic principles. This is especially so considering the millions of tons of leaf waste incinerated or put into major greenhouse gas emitting landfills annually [9].

Another application of lignocellulose leaf skeletons is in utilizing them as free-standing, metallized flexible electrodes in sensors/transducers, or as quasi-transparent electrodes for photon flux applications such as solar cells, OLEDs etc. [10] [11] [12] [13]. This is possible owing to their high transparency and flexibility even when embedded with thin layers of metals such as Cu or Ag. This opens doors for the ingress of naturally occurring, partially treated materials as potential replacements of expensive and process-intensive ones like Indium Tin Oxide (ITO) without reducing the state-of-the-art performance metrics or efficiencies in some applications [14] [15].

Here, the leaves of the Magnolia tree are chosen as a suitable source for extracting viable leaf skeletons owing to their thick leaf structures and dense venation [16]. Freshly plucked leaves were cut to 2.5 cm² samples and submerged in concentrated $Na_2CO_3$ solution prior to heating at 90°C for 5 hours followed by ultrasonication. The process was repeated until the leaf skeletons were fully exposed. The skeletons were subsequently washed in 5% $H_2O_2$ and dried in an oven at 60°C overnight under a flat load.

## II. Results and discussion

Ethyl cellulose (EC) dissolved in 2-Butoxyethanol (8 g in 36 ml) is utilised as a high viscosity solution to coat bare leaf skeletons (LS) via facile dip-coating. The freshly dip-coated samples are dried in open air at 60°C in a custom-built vertical-alignment hot air dryer to create non-porous substrates as shown in Fig. 1.

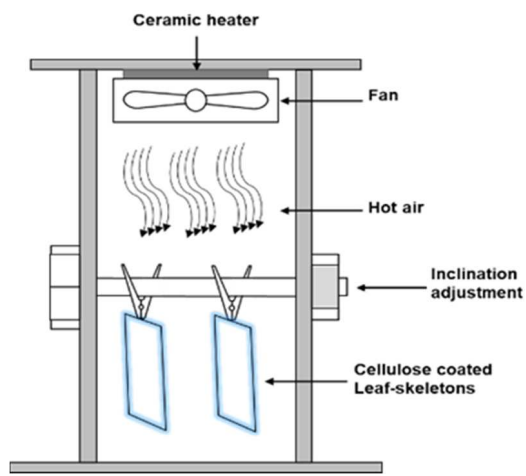

**Figure 1.** Vertical curing set-up for dip-coated substrates

### A. Leaf skeletons as biodegradable substrates

The structures when dip-coated in EC solution allow for the EC molecules to fill up the vacancies in the porous, quasi-fractal geometry of the leaf skeletons. This results in a substrate (now referred to as LS-EC) with emergent properties because of the structural stability provided by the lignocellulose skeleton and the elasticity provided by the cellulose polymer chains.

Testing showed that the substrate has good chemical resistance (EC is stable under pH ranges of 3 to 11) [17] and high hydrophobicity [18]. The substrate additionally remains unaffected at temperature ranges below 150°C which makes the LS-EC substrate suitable for additive fabrication techniques such as functional printing where ink drying temperatures between 90°C and 120°C are common.

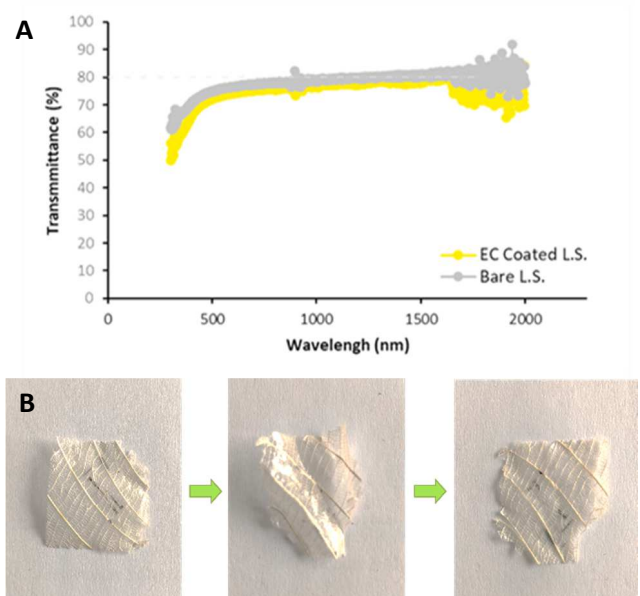

**Figure 2.** (A) Optical transmittance characteristics of an uncoated leaf skeleton (upper) and an EC coated leaf skeleton (lower) (B) Crush test performed on the LS-EC substrate showing pristine (1), crushed (2), and subsequently recovered (3) substrate.

Measurement of the optical transmittance properties of the EC-LS substrates before and after substrate formation (Fig. 2A) show that impregnation of the pores with EC does not change the transparency of the original material significantly and the light transmittance in the visible spectrum lies around 75%.

Fig. 2B demonstrates the ability of the substrate to recover from excessive bending and even crushing. The pristine substrate shown in Fig.3A.1 was manually crushed until multiple creases were evident on the 2.5 x 2.5 cm$^2$ sample with audible breakage of the internal lignocellulose structures and some physical damage around the corners (Fig. 2B.2). The substrate recovered its original flat shape without any punctures when heated at 90°C for 30 to 60 seconds with a flat load on top to relieve stress laterally (Fig. 2B. 3).

### B. Leaf skeletons as conducting electrodes

After the lignocellulose leaf skeletons were successfully utilised for substrate fabrication via impregnation with EC, their quasi transparency and quasi-fractal nature was also utilised for the creation of transparent conducting electrodes (TCE).

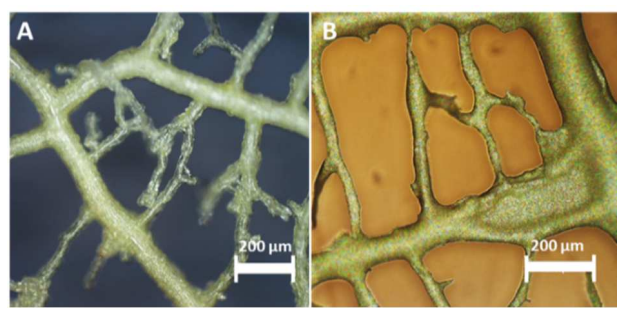

**Figure 3.** (A) Uncoated leaf skeleton (200 μm scale bar) (B) Ag coated leaf skeleton (200 μm scale bar).

Pretreatment was performed on the leaf skeletons by soaking them in 10 mM solution of Tri-dodecylmethylammonium chloride (TDMAC) for 5 minutes at 90°C prior to dip-coating in microparticle based Ag ink. This resulted in the partial electrostatic binding of Ag particles to the LS. The sheet resistance measurement results are shown in Fig. 4 where the sheet resistance of the TDMAC pre-treated LS is an order of magnitude lower than that of samples simply dip-coated with no chemical pre-treatment.

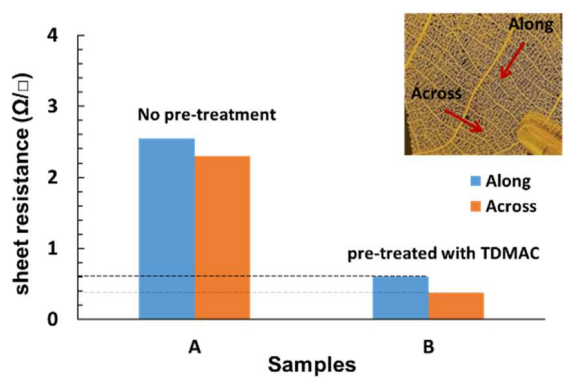

**Figure 4.** Change in sheet resistance upon dip-coating with Ag - without chemical pre-treatment (A) and with TDMAC pre-treatment (B).

In order to avoid erroneous measurements caused from agglomerated Ag on the secondary lignocellulose vein structures (Fig. 4 inset) sheet resistance measurements were performed across as-well-as along the 2.5 cm² LS samples.

Electro-thermal stability of the samples was measured by passing DC current through the Ag coated LS (henceforth referred to as Ag-LS). The current from a constant current source was increased until a value of 4 Amperes was reached. It was noted that the samples could handle such current ranges indefinitely till a value of 6.2 Amperes. Fig. 5A shows that even at 4 Amperes of constant current flow, Ag-LS electrodes could dissipate the heat due to their quasi-fractal structure and maintain a surface temperature of around 60°C.

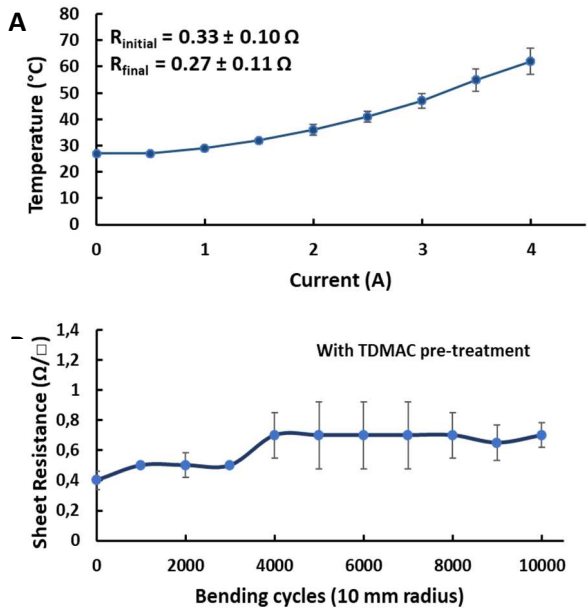

**Figure 5. (A)** Rise in temperature upon DC current flow **(B)** Change in sheet resistance upon 10,000 bend-relax cycles.

Mechanical stability of the Ag-LS was tested on a custom bending machine. Fig. 5(C) shows that the coating of Ag on the LS using TDMAC results in a robust bonding between Ag particles with LS resulting in a negligible change in sheet resistance even after 10,000 bending cycles.

The two most important factors in electrode implementation in photon flux applications like organic photovoltaics are the optical transmission (transmittance) value and the sheet resistance of the transparent electrodes. As a figure of merit, the minimum required transmittance value for suitability in photon flux applications is around 80% while the sheet resistance values can vary from < 1 Ω/□ to > 200 Ω/□ [19]. Although good conductivity values are achieved with the TDMAS pre-treatment, light transmittance for the Ag-LS electrodes also needs to be improved. Cu electroplating is implemented for this purpose.

The sheet resistance after dip-coating in highly diluted (~ 10 Pa·s) Ag ink was found to be 72 Ω/□ due to very little Ag adhesion. This also resulted in an optical transmittance of around 80% (Fig. 6A) because of lower surface coverage. This thin Ag coat was used as a seed layer for the deposition of atomic Cu via electroplating. Fig. 6(B) shows that electroplating can be performed until the desired conductivity is reached (final sheet resistance = 0.71 Ω/□) without affecting the high optical transmittance value significantly. With sheet resistances lowered via Cu electroplating and transmittances approaching 80%, the results presented here can rival some state-of-the-art properties of much more energy intensive and complex techniques for fabricating transparent electrodes [19].

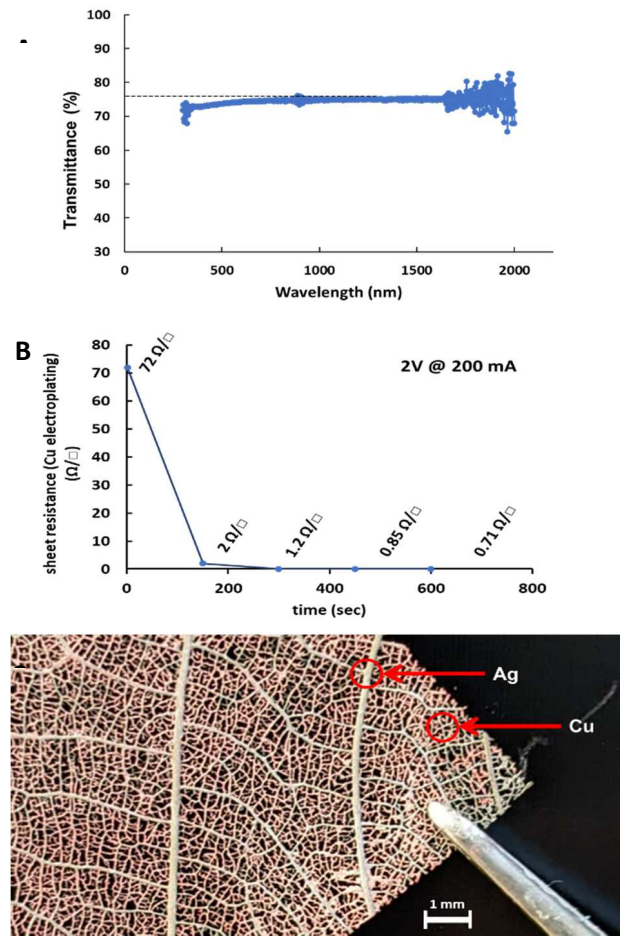

**Figure 6. (A)** Optical transmittance of copper electroplated Ag-LS **(B)** Sheet resistance of Ag-LS at different times during Cu electroplating **(C)** Cu coat after electroplating (Scale bar 1 mm).

## III. CONCLUSION

Lignocellulose leaf skeleton structures extracted from tree leaves were incorporated in the formation of flexible bio-sourced substrates as well as for the creation of quasi-transparent, metallized electrodes. Additionally, a methodology was developed to independently tune the optical transmittance and the sheet resistance of the electrodes. The low-cost fabrication and ease of processing makes these ubiquitously available biomaterials interesting in terms of their properties as substrates as well as flexible, light-admitting electrodes, especially in the field of organic electronics.

ACKNOWLEDGMENT

The authors are thankful to the European Union for funding the project EMERGE. The EMERGE project has received funding from the European Union's Horizon 2020 research and Innovation programme under grant agreement no. 101008701. H.K. are grateful for funding from the German Research Foundation (DFG) under the grant KL 2961/5-1.